\documentclass{caosp}

\usepackage{graphicx}
\usepackage{amssymb}

\articleNo{123}
\pubyear{2005}
\volume{35}
\volnumber{3}
\firstpage{1}
\received{May 1, 2005}
\accepted{August 28, 2005}

\begin{document}

%
\htitle{The Fly's Eye project: sidereal tracking on a hexapod mount}
\hauthor{K. Vida {\it et al.}}

\title{The Fly's Eye project}

\subtitle{Sidereal tracking on a hexapod mount}

%
\author{
	Kriszti\'an Vida \inst{1} \and
	Andr\'as P\'al \inst{1,} \inst{2} \and
	L\'aszl\'o~M\'esz\'aros \inst{1,} \inst{2} \and
	Gergely~Cs\'ep\'any \inst{1,} \inst{2} \and
	Attila Jask\'o \inst{1,} \inst{3} \and 
	Gy\"orgy Mez\H{o} \inst{1} \and
	Katalin~Ol\'ah \inst{1}
       }

%
\institute{	MTA Research Centre for Astronomy and Earth Sciences, 
	        Konkoly Thege Mikl\'os \'ut 15-17,
        	Budapest H-1121, Hungary \email{kvida@flyseye.net}
	         \and 
		Department of Astronomy, Lor\'and E\"otv\"os University,
	       	P\'azm\'any P. st. 1/A,
	        Budapest H-1117, Hungary 
		 \and
		Budapest University of Technology and Economics,
		M\H{u}egyetem rkp. 3., 
		Budapest H-1111, Hungary 
          }

\date{March 8, 2003}

\maketitle


\begin{abstract}
The driving objective of the Fly's Eye Project is a high resolution, 
high coverage
time-domain survey in multiple optical passbands: our goal is to
cover the entire visible sky above the 30$^\circ$ horizontal altitude
with a cadence of ${\sim 3\,\rm min}$. Imaging is intended to perform 
with 19 wide-field cameras mounted on a hexapod platform. The essence of the 
hexapod allows us to build an instrument that does not require any kind 
of precise alignment and, in addition, the similar mechanics can be 
involved independently from the geographical location of the device.
Here we summarize our early results with a single camera, 
focusing on the sidereal tracking as it is performed 
with the hexapod built by our group. 
\end{abstract}


\section{Introduction}
\label{sec:introduction}

In the recent years, many initiatives have been started 
in order to perform optical astronomical surveys in the time domain. 
Some of the projects intend to focus on a few specific discipline,
e.g. the \emph{Kepler} mission (on exoplanetary and asteroseismology research, 
see e.g. {Borucki et al.~2007}) or cover dozens of independent scientific
key projects, such as the Pan-STARRS ({Kaiser et al.~2002}) or 
the Large Synoptic Survey Telescope (LSST, {Ivezi\'c et al.~2008}). 
These surveys attain their success due to the extreme light collecting
power or \emph{\'etendue}, which is the multiple of the net aperture area
and the effective field-of-view of the imaging optics (basically a measure of light collecting power). 
The aforementioned projects perform observations by either covering a smaller celestial area
with frequent sampling (such as \emph{Kepler}) or a larger areas with
sparser sampling (Pan-STARSS or LSST), however, all of these deal with
high imaging resolution. The goal of the Fly's Eye Project is do develop, and
operate a high coverage, high cadence, but lower imaging resolution 
instrumentation with a comparable \'etendue to the previously mentioned 
projects The scientific goals 
of our project also cover dozens of astrophysical phenomena,
as it is described in {P\'al et al. (2013)}. Our expectation is to achieve
photometric precision in the millimagnitude level for stars with
a brightness of Sloan $r=10^{\rm m}$ as well as a faint depth of $r\lesssim 15^{\rm m}$
with S/N of 5 or more for isolated objects.

The high coverage and high cadence is attained by observing the 
visible sky simultaneously using numerous wide-field cameras
(similarly to Deeg et al. 2004 or Pepper et al. 2007). Although
the imaging resolution is essentially low ($22^{\prime\prime}/{\rm px}$), even
high cadence images require precise sidereal tracking during the exposures.
In our design, this tracking is achieved by a hexapod mount (also known 
as Steward-platform), which has many advantages due to conventional
bi-axis mechanisms, including the lack of proper adjustment and the
usability of the same instrument independently from the geographical location.
In this paper we briefly summarize the key concepts of the hexapod 
itself as well as the results of our first tests related to the sidereal
tracking. 

\section{The hexapod design}
\label{sec:hexapod}

Due to its complexity, hexapods are barely involved as a 
primary mount in astrophysical applications. Mainly, these are used
as a support for secondary mirrors (see e.g. {Geijo et al.~2006}),
and there are direct applications in the field of radio astronomy
({Koch et al. 2009}) as well as optical spectroscopy ({Chini 2000}).
Since with the exception of the Fly's Eye initiative, 
there is no direct application for optical imaging, 
in this section, we summarize the key concepts of our hexapod design.

\paragraph{Mechanics.}

As its name suggests, our hexapod involves six identical linear actuators
with two universal joints mounted at both ends. Our choice of electromechanical
actuators exploit a jack screw in order to transform rotary motion 
into linear one. The total stroke of our actuator is limited to $100\,{\rm mm}$
while the net length of a single leg in ``home position'' is approximately
$510\,{\rm mm}$. The net length is defined as the distance between the
two corresponding universal joint center points while the ``home position''
is defined at halfway between the fully retracted and full stroke state. 
Magnetomechanical switches ensure the proper limiting of each actuator
while the position feedback is done by a Hall effect based rotary encoder
having a resolution equivalent to $\pm0.1\,{\mu m}$ stroke. 

\begin{figure}[!t]
\centering
\includegraphics[width=55mm]{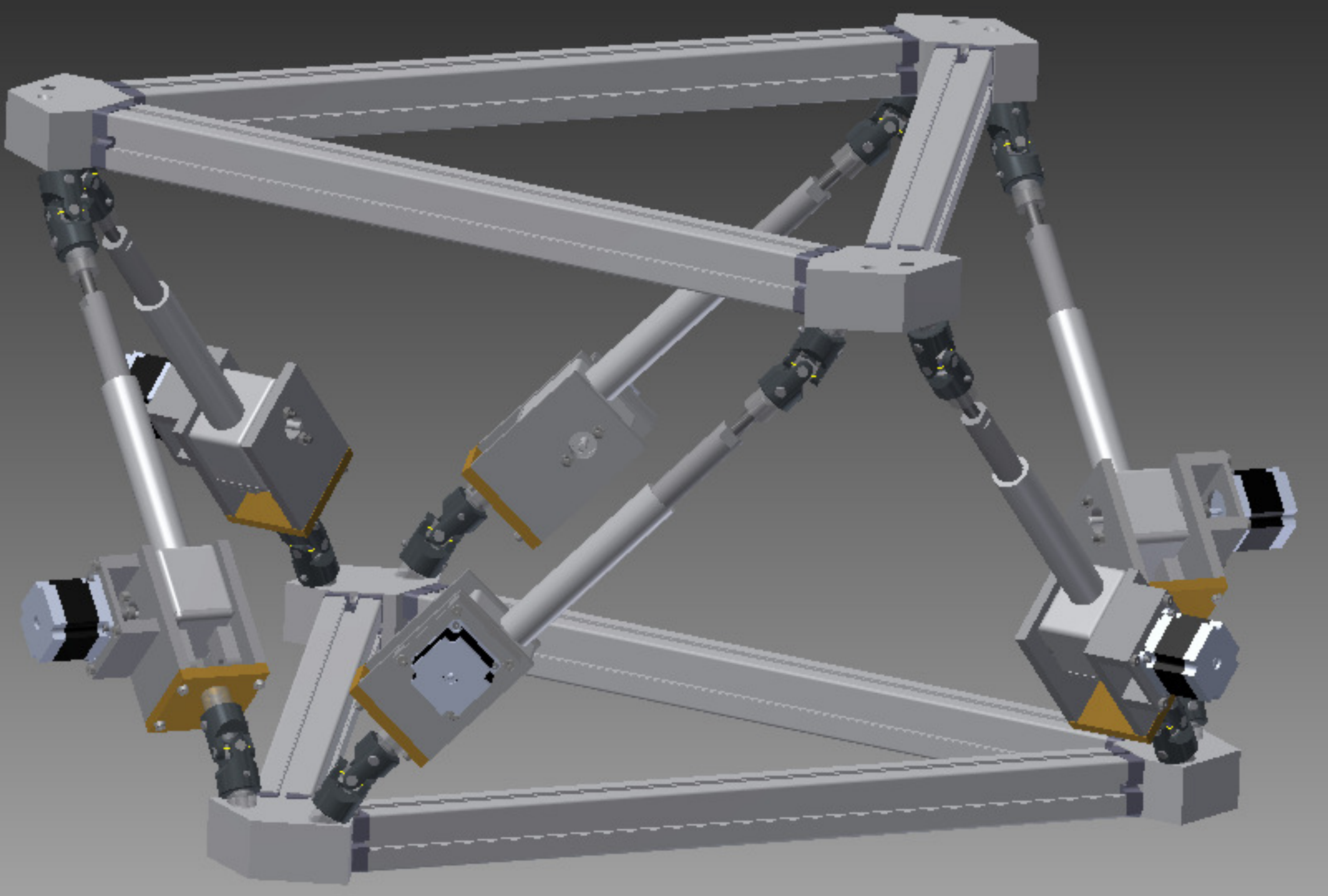}
\includegraphics[width=55mm]{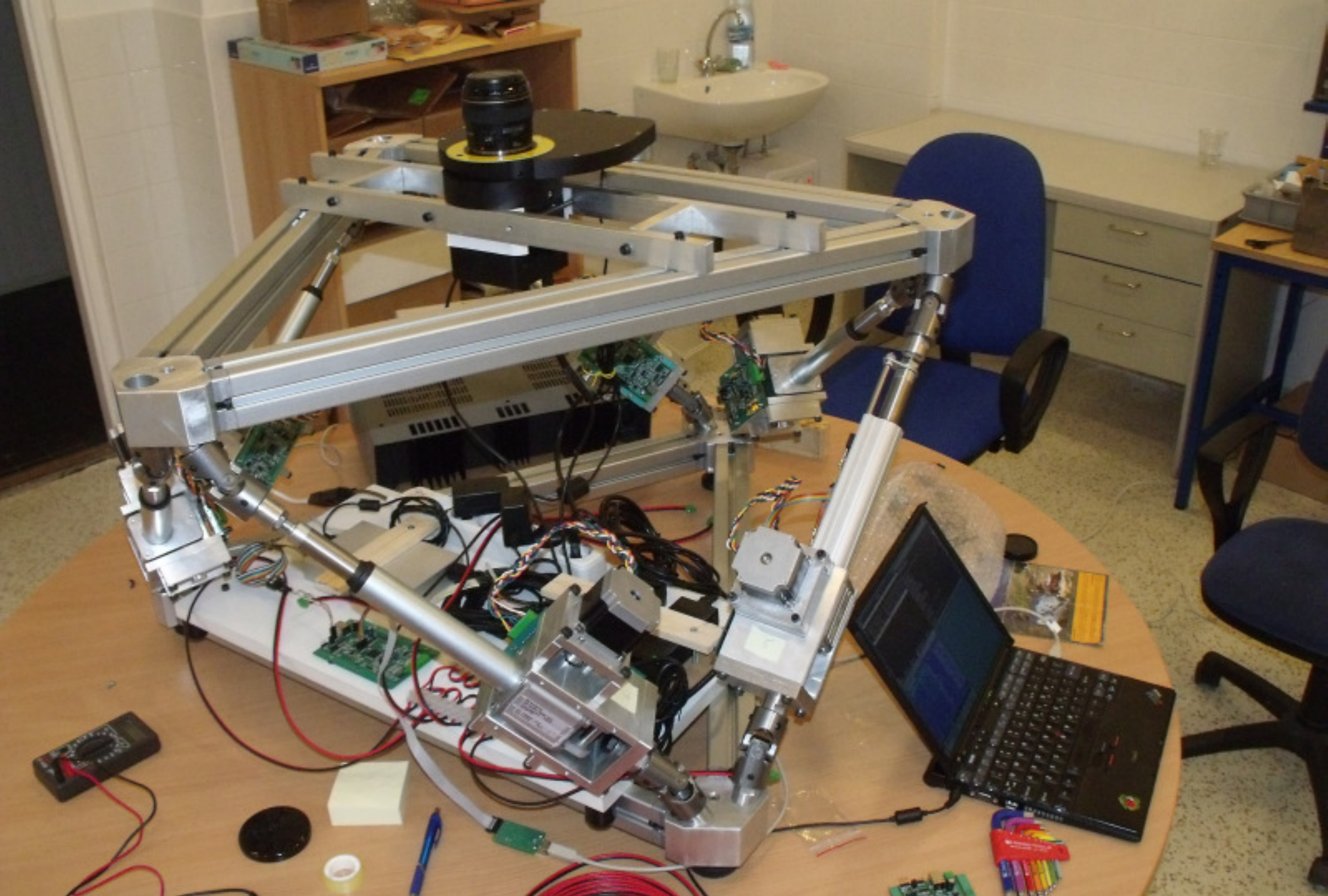}
\caption{Left panel: a CAD view of the hexapod skeleton, showing the base and
payload platforms and the six legs. Right panel: the fully assembled 
hexapod with a single camera--filter--lens set in the lab, just prior 
delivery for first light tests. 
\label{fig:hexapods}}
\end{figure}

Both on the base and the payload platform, the universal joints are mounted
to a triangle-shaped structure whose side is approximately $700\,{\rm mm}$.
Hence, the characteristic instrument size and the actuator travel length
($100\,{\rm mm}$) yield us approximately $\pm 10^\circ$
of rotary domain of the hexapod, that is sufficient for more than an hour
of sidereal tracking. The CAD view as well as the complete hexapod
(with a camera and optics at the current state) can be seen in Fig.~\ref{fig:hexapods}.

%

\paragraph{Electronics and firmware.}

The actuators are driven by stepper motors that are controlled by
a customized electronics board, one mounted on each leg. Each of the boards
is connected to the Hall rotary encoder electronics that, with additional
non-volatile ferroelectric RAM based storage, allows a 
complete stateless operation of the full device. The core of each board is 
an AVR architecture microcontroller with identical firmwares. 
The real-time code multiplexes the communication
with the control bus (see below), the motor driver circuit and the 
position encoder.

\paragraph{Control subsystem and software.}
\label{sec:hexapod:software}

The onboard leg electronics are connected with each other by an RS485 bus.
This bus system allows a complete synchronized operation: the leg 
movement parameters (duration, stroke/retract, higher order polynomial 
coefficients etc.) are uploaded in a unicast manner to each leg while 
the ``start'' command is a broadcast message. During motion, the status
of the legs can  be polled directly.
Although RS485 lets distant parts to be connected, this bus is also driven
locally by a single-board computer (SBC) which serializes commands
received on TCP/IP to the RS485 bus. 

\paragraph{Camera control and data acquisition.}

During the test runs, we employed a single CCD camera with a KAF-16803
detector and $f/1.8$, $f=85\,{\rm mm}$ lens, equipped 
with $50\times 50\,{\rm mm}$ Sloan g', r' and i' filters. The 
camera has been mounted on the geometric center of the hexapod
payload platform (see Fig.~\ref{fig:hexapods}., right panel)
The camera and the filter wheel are controlled via USB while the
USB is hosted on the same SBC that drives the RS485 bus (see above). The
electric lens focusing is realized via an SPI (Serial Peripheral Interface) bus, also hosted on a
RS485 node, connected to the same bus.
Hence, the traffic of the
whole device, including the camera, filter, hexapod mount control, as
well as other housekeeping sensors (humidity, temperature, etc.) are
tunnelled via multiple TCP/IP channels. The data acquisition could 
therefore be controlled by any machine connected to the Internet.
In our tests, while the device was located at Piszk\'estet\H{o} station, in most of
the time, data acquisition was managed by a computer located at Budapest. 

\section{Sidereal tracking}
\label{sec:siderealtracking}

During our first test runs, we acquired more than 4,000 individual frames
in a filter sequence of g'-r'-i'-r'. For simplicity, the subsequent data acquisition 
steps (hexapod repositioning between exposures, filter changing, focusing,
readout) were not done in parallel, hence the duty cycle is currently
somewhat smaller than our final goal (70\% instead of 90\%+). 

\begin{figure}[!t]
\centerline{%
\includegraphics[width=18mm]{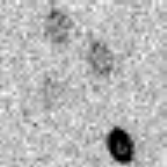}\,
\includegraphics[width=18mm]{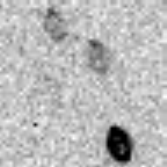}\,
\includegraphics[width=18mm]{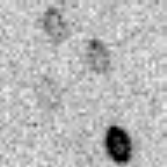}\,
\includegraphics[width=18mm]{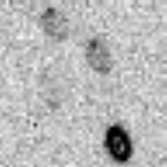}\,
\includegraphics[width=18mm]{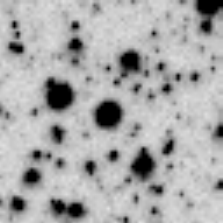}\,
\includegraphics[width=18mm]{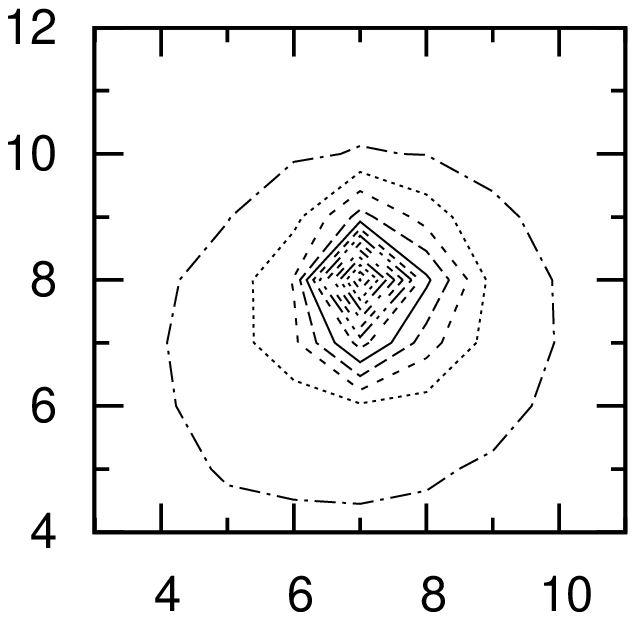}%
}
\caption{Left four panels: Tracking using 
an $f=800\,{\rm mm}$ lens during a 3\,min interval (note: the stellar
profiles are defocused on these $1.8'\times 1.8'$ stamps). Second to the right: Image stamp of 
$64\times 64$\, pixels, taken with an $f=85\,{\rm mm}$ lens, exposure time: 
130 seconds. Right panel: PSF of the stellar profile at the 
center of the previous image. 
\label{fig:stamps}}
\end{figure}

\subsection{Self-calibration}

It is easy to see that 3 linear combinations of the 6 leg strokes
yields nearly pure pitch, roll or yaw rotations. By setting properly
the actuator stroke speeds using this 3 combinations, one is able to attain
proper sidereal tracking with the hexapod. As a self-calibration
procedure, we gathered 4 series of subsequent image pairs in order to 
obtain the numerical derivative of the field centroid coordinates with
respect to the pitch, roll and yaw speed offsets. This is done under
the assumption that the device is exactly aligned to 
the compass points and to the horizon. For this analysis, we employed
the tasks of the FITSH package (P\'al, 2012).

Since the hexapod geometry (determined by the 
universal joint center positions) is known with a significantly 
better relative accuracy ($\approx 3\times 10^{-4}$, assuming 
an assembly precision of $\approx 0.3{\rm mm}$) than the alignment
($\approx 1 \dots 5 \times 10^{-2}$), we could set up a set of equations
by saying that 1) the rotation speed must be equal to the sidereal 
and 2+3) the apparent drift of the centroid (on the CCD plane) 
must be zero. By solving 
these 3 equations, we could determine the offsets that should
be added to the pitch, roll and yaw speeds.

\subsection{Precision and repeatability}

Our initial results show that despite of its simplicity, the above 
procedure yields indeed proper sidereal tracking on the time scale
of few minutes, even considering that actuator stroke speeds 
are constants on the timescale of exposures. 
We performed these tests not only with the $f=85\,{\rm mm}$
lens but also with a catadioptric optics of $f/8$, $f=800\,{\rm mm}$
($\approx 2.3^{\prime\prime}/{\rm px}$, see also Fig.~\ref{fig:stamps}). 
The latter tests yielded a tracking 
drift of $0.5^{\prime\prime}{\rm min}^{-1}$ (see left 
four panels of Fig.~\ref{fig:stamps}) which is equivalent
to a relative error of $6\times 10^{-4}$. This value is well comparable to the
\emph{a priori} assumption of the assembly precision.

As it is discussed in {P\'al et al. (2013)}, images are acquired by
synchronization to Greenwich sidereal time. By comparing
field centroid celestial coordinates of corresponding images
taken on subsequent nights, we could reliably characterize the repeatability
of the instrument since the hexapod performs hundreds of independent 
movements between two respective frames. We found that this repeatability
is in the level of some tenths of pixels using the $f=85\,{\rm mm}$ optics,
that is in the range of a few arcseconds. 
 
\section{Summary}
\label{sec:summary}

This paper described the first results of the Fly's Eye project
related to the sidereal tracking as it is performed with a hexapod mount.
Despite of the very simple tracking algorithm and self-calibration procedure,
we found that the precision is well beyond within a magnitude than our needs,
justifying that the hexapod is an adequate mount for such an optical imaging
instrument. The complete analysis of the test runs will be presented
in further paper(s).


\acknowledgements
The ``Fly's Eye'' project is supported by the Hungarian Academy of
Sciences via the grant LP2012-31. 
K. V. and O. K. acknowledge the support of the OTKA-K81421 grant.
This work was supported by the HUMAN MB08C 81013 grant of the MAG Zrt.
We thank
H. Deeg (PI of the PASS project) for the useful discussions.
We also thank to our colleagues F. Schlaffer, E. Farkas and L.~D\"obrentei
for their help during the hexapod development.

{}


\begin{thebibliography}{}
\bibitem[Borucki et al. (2007)]{borucki2007}
Borucki, W. J.~et al.:
2007, ASP Conf.~Ser., {\bf 366}, 309

\bibitem[Chini (2000)]{chini2000}
Chini, R.: 
2000, {\it Rev. Mod. Astron.} {\bf 13}, 257

\bibitem[Deeg et al. (2004)]{deeg2004}
Deeg, H. J et al.:
2004, {\it \pasp} {\bf 116}, 985

\bibitem[Geijo et al. (2006)]{geijo2006}
Geijo, E. M. et al.:
2006, {\it Proc.~SPIE} {\bf 6273}, 99

\bibitem[Koch et al. (2009)]{koch2009}
Koch, P. M. et al.:
2009, {\it \apj} {\bf 694}, 1670

\bibitem[Ivezi\'c et al. (2008)]{ivezic2008}
Ivezi\'c, \v{Z}. et al.:
2008, arXiv:0805.2366

\bibitem[Kaiser et al. (2002)]{kaiser2002}
Kaiser, N. et al.:
2002, {\it Proc.~SPIE} {\bf 4836}, 154

\bibitem[P\'al (2012)]{pal2012}
P\'al, A.:
2012, {\it \mnras} {\bf 421}, 1825

\bibitem[P\'al et al. (2013)]{pal2013}
P\'al, A. et al.
2013: {\it \an} {\bf 334}, 932 

\bibitem[Pepper et al. (2007)]{pepper2007}
Pepper, J. et al.:
2007, {\it \pasp} {\bf 119}, 923

\end{thebibliography}
\end{document}